\documentclass[twocolumn,tighten]{aastex63}

\usepackage{times,natbib,graphicx,amsmath,multirow, xspace}
\usepackage{xcolor}
\usepackage{appendix}
\usepackage{lineno}

\newcommand{\nustar}{NuSTAR\xspace}

\newcommand{\nicer}{NICER\xspace}

\newcommand{\lumcgs}{ergs~s$^{-1}$\xspace}
\newcommand{\rin}{$R_{\rm in}$\xspace}
\newcommand{\rg}{$R_{g}$\xspace}
\newcommand{\risco}{$R_{\mathrm{ISCO}}$\xspace}

\newcommand{\relxill}{{\sc relxill}\xspace}
\newcommand{\relxillns}{{\sc relxillNS}\xspace}

\newcommand{\xillver}{{\sc xillver}\xspace}
\newcommand{\xillverco}{{\sc xillverCO}\xspace}
\newcommand{\source}{SLX~1735$-$269\xspace}
\DeclareUnicodeCharacter{2212}{-}
\definecolor{pink}{RGB}{212,76,133}
\definecolor{sky}{RGB}{51,160,232}
\definecolor{lavender}{RGB}{170,120,240}

\shorttitle{NICER-NuSTAR View of SLX 1735-269}
\shortauthors{Moutard et al.}

\begin{document}

\title{Investigating the Ultra-Compact X-ray Binary Candidate \source with \nicer and \nustar}
\correspondingauthor{D. L. Moutard}
\email{david.moutard@wayne.edu}

\author{D.~L.~Moutard}
\author[0000-0002-8961-939X]{R.~M.~Ludlam}
\author[0000-0003-0440-7978]{M.~Sudha}
\affiliation{Department of Physics \& Astronomy, Wayne State University, 666 West Hancock Street, Detroit, MI 48201, USA}

\author{D.~J.~K.~Buisson}
\affiliation{Independent Researcher, UK}
\author[0000-0002-8294-9281]{E.~M.~Cackett}
\affiliation{Department of Physics \& Astronomy, Wayne State University, 666 West Hancock Street, Detroit, MI 48201, USA}
\author[0000-0002-0092-3548]{N.~Degenaar}
\affiliation{Anton Pannekoek Institute for Astronomy, University of Amsterdam, Science Park 904, 1098 XH Amsterdam, Netherlands}
\author[0000-0002-9378-4072]{A.~C.~Fabian}
\affiliation{Institute of Astronomy, University of Cambridge, Madingley Rd, Cambridge CB3 0HA, UK}
\author[0000-0003-3105-2615]{P.~Gandhi}
\affiliation{School of Physics and Astronomy, University of Southampton, University Road, Southampton SO17 1BJ, UK}
\author[0000-0003-3828-2448]{J.~A.~Garc\'{i}a}
\affiliation{Cahill Center for Astronomy and Astrophysics, California Institute of Technology, 1200 E. California Blvd, MC 290-17, Pasadena, CA, 91125, USA}
\author[0000-0001-6715-0423]{A.~W.~Shaw}
\affiliation{Department of Physics and Astronomy, Butler University, Indianapolis, IN, 46208, USA}
\author[0000-0001-5506-9855]{J.~A.~Tomsick}
\affiliation{Space Sciences Lab, University of California, Berkeley, 7 Gauss Way, Berkeley, CA, 94720, USA}

\begin{abstract}
We present two simultaneous \nicer and \nustar observations of the ultra-compact X-ray binary (UCXB) candidate \source\ while the source was in two different spectral states. Using various reflection modeling techniques, we find that \xillverco, a model used for fitting X-ray spectra of UCXBs with high carbon and oxygen abundances is an improvement over \relxill or \relxillns, which instead contains solar-like chemical abundances. This provides indirect evidence in support of the source being ultra-compact. We also use this reflection model to get a preliminary measurement of the inclination of the system, {\it i} $= 57^{+23}_{-7}$ degrees. This is consistent with our timing analysis, where a lack of eclipses indicates an inclination of $i<80^{\circ}$. The timing analysis is otherwise inconclusive, and we can not confidently measure the orbital period of the system.\\ 
\end{abstract}

\section{Introduction}
A low-mass X-ray binary (LMXB) is a system comprised of a compact object, a neutron star (NS) or black hole (BH), interacting gravitationally with a main sequence, sub-giant, or red giant star (which we may call \textit{canonical} LMXBs). In these systems, the companion star fills its Roche Lobe, and then deposits matter into an accretion disk surrounding the compact object. An ultra-compact X-ray binary (UCXB) is a subclass of LMXB differentiated by a much shorter orbital period, generally defined to be $<$80 minutes, compared to the typical periods of hours to days that are seen in LMXBs \citep{bahramian23}. This shorter period is caused by a more compact companion than a main sequence star, such as a white dwarf (WD) or helium star \citep{nelson86, savonije86}. These companions have a notably different chemical composition than their main sequence or red giant counterparts, often lacking hydrogen and helium, and containing an overabundance of carbon and oxygen. LMXBs are well studied systems, used to understand generally accretion physics and the physics of compact objects and in the era of multi-messenger astronomy they can be considered as a source of gravitational waves \citep{chen20}. 

In LMXB and UCXB systems it is believed that the X-rays originate from the region of closest accretion inflow, where material transitions from the accretion disk to falling onto the compact object. Near the compact object we expect an X-ray corona- a source of non-thermal photons generated from the Compton up-scattering of seed photons from the accretion disk, or in the case of a NS LMXB, perhaps a boundary layer of material that surrounds the surface of the NS \citep{syunyaev91}. Some of these hard X-rays should reach the observer directly, but we also expect to observe the interaction between these photons and the rest of the LMXB system. This can manifest as reflection features, where coronal X-rays scatter off the disk, and are then reprocessed. A common feature of this reflection is the Fe K$\alpha$ line around 6.4 keV, but the unique composition of UCXBs means that we may also see an \ion{O}{8} Ly$\alpha$ feature at around 0.67 keV. In UCXBs we also sometimes see a suppression of the Fe K$\alpha$ line \citep{koliopanos14}. These reflected features experience a relativistic broadening, as the disk material orbits rapidly around the compact object. In X-ray studies, these features are believed to arise from the region of the disk nearest the compact object \citep{fabian89}. Because of this, we can use the broadening of the reflected emission to determine the radius at which the innermost region of the disk sits. For NS LMXBs, this can provide an upper bound on the radius of the NS, which is important for understanding the NS equation of state \citep{cackett08,miller13,ludlam17}. For a recent comprehensive review of reflection studies in NS LMXBs, see \cite{ludlam24}. In reflection studies we are therefore able to model the spectral contribution from 3-4 different components: non-thermal photons from the corona, thermal photons from the disk, reflected emission, and/or thermal emission from the NS and boundary layer itself.

\source was discovered in 1985 during the Spacelab 2 mission  during X-ray observations of the Galactic center \citep{skinner87}. The existence of thermonuclear X-ray bursts \citep{bazzano97} as well as the spectral shape \citep{david97} demonstrates that the compact object in this system is a NS, but the companion is still poorly understood. \cite{intzand07} proposes the UCXB candidacy based on its low luminosities and the frequency of bursts. It has been found that almost all UCXBs occupy the lowest accretion rate regimes, and accretion rate is directly proportional to the luminosity. An increased burst recurrence time can be explained by a lower accretion rate. \cite{molkov05} detects long bursts in this system as well, up to $\sim$ 2ks. This is explained to likely be the burning of a mixed pile of hydrogen and helium. This helium burning scenario has been used to explain long bursts in UCXBs in the past \citep{cumming06}. However, lack of optical spectra to confirm the presence of carbon or oxygen lines and no studies of the orbital period means we can not verify the UCXB nature of \source. The source is localized with subarcsecond accuracy in Chandra at Galactic coordinates $\ell = 0.796$, $b = 2.400$ \citep{wilson03}. A possible optical counterpart that is spatially coincident with the X-ray source exists near the edge of the Chandra positional uncertainty for \source, though it has not been spectrally identified. The counterpart also exhibits a shape which can not be ruled out as having a double nature, so estimates in position and magnitude have additional uncertainty \citep{zolotukhin11}. Because \source is so close to the Galactic center, the high column density of neutral hydrogen may make optical studies to search for key UCXB features difficult. In this paper we use simultaneous \nicer and \nustar observations with reflection modeling techniques to better understand the source. In Section \ref{sec:observations} we discuss the details of the data reduction and observations and in Section \ref{sec:results} we show the results of our spectral analysis. In Section \ref{sec:discussion} we discuss the implications of these results, then summarize the results and conclude the paper.

\section{Observations and Data Reduction}\label{sec:observations}
\begin{figure}[t!]
\begin{center}
\includegraphics[width=0.48\textwidth, trim = 50 0 23 0, clip]{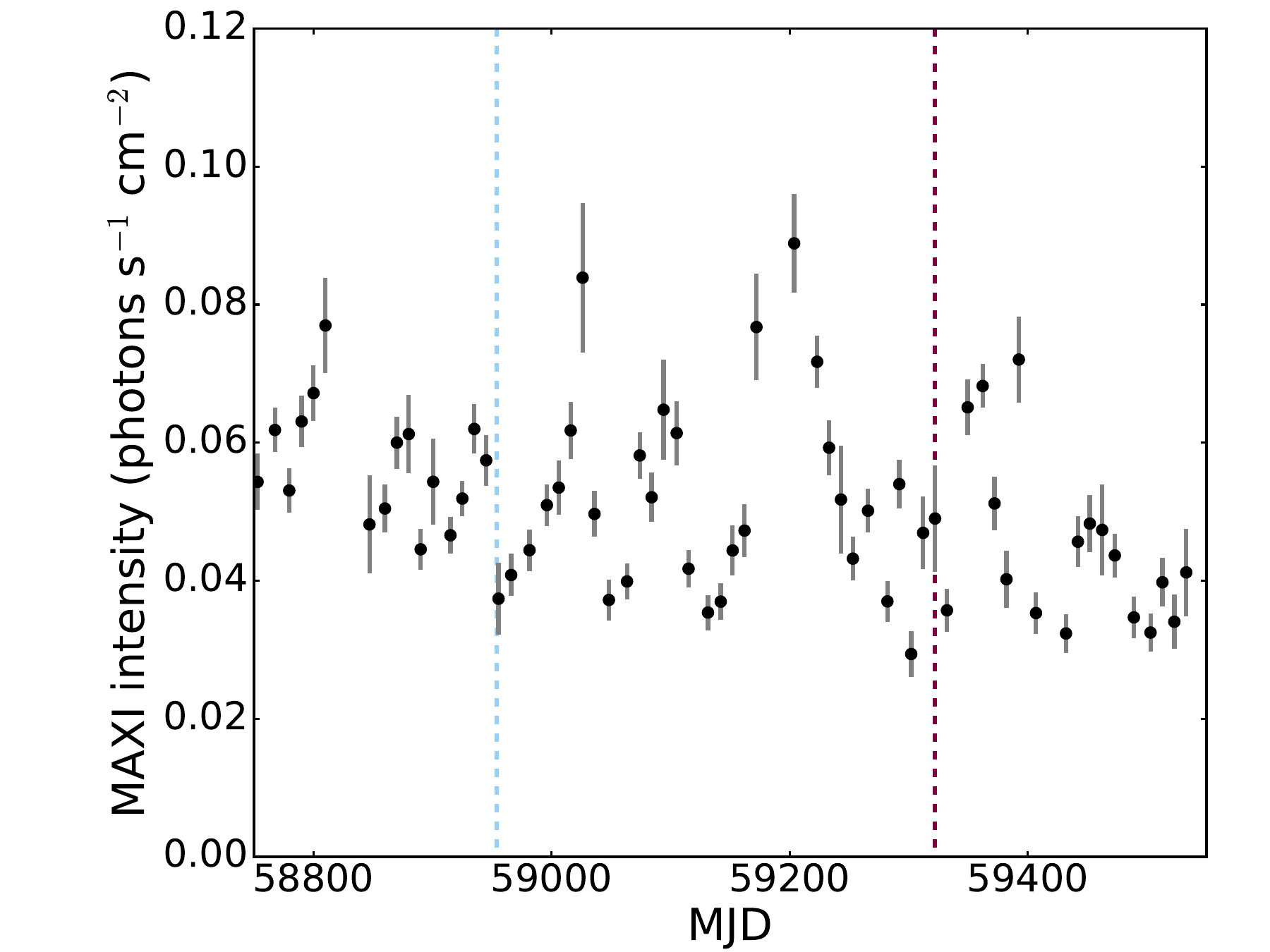}
\caption{A long term MAXI light curve for the source. Vertical lines indicate the dates at which Obs 1 (left, blue) and Obs 2 (right, red) occur. We see that Obs 1 occurs during a lower flux state than Obs 2. The MAXI data are binned to 10 days.}
\label{fig:maxi}
\end{center}
\end{figure}

\source was observed on two separate occasions roughly one year apart with both \nicer and \nustar simultaneously. More detailed information about these observations can be seen in Table \ref{tab:obs}. We reduce the \nustar data using {\sc nustardas} v2.1.2 and {\sc caldb} 20230816. The light curves and spectra were extracted using regions with a diameter of 100$''$ centered on the source. The backgrounds were also extracted using 100$''$ apertures, but centered elsewhere. Obs 2 displayed some contamination from stray light, but these do not overlap the source itself, and backgrounds were selected such that they did not contain this stray light contamination. The \nicer data were calibrated using {\sc nicerdas} 2023-08-22\_v011a and {\sc caldb} 20221001. This calibration was done first by the use of {\sc nicerl2} for geomagnetic prefiltering. The {\sc nimaketime} command was used to generate good time intervals (GTIs) with low particle background (KP $<$ 5). Other cuts are made to eliminate particle overshoots ({\sc cor\_range} 1.5-20 and {\sc overonly\_range} 0-2). Then {\sc nicerl3-spect} is used to create NICER spectra, background, and response files and {\sc nicerl3-lc} is used to generate light curves. These instances of {\sc nicerl3-spect} and {\sc nicerl3-lc} utilize the 3C50 background model \citep{remillard22}. Figure \ref{fig:maxi} shows the time of these observations on a MAXI light curve.

\begin{table}[t!]
\caption{\source Observation Information}
\label{tab:obs}
\begin{center}
\begin{tabular}{llcccc}
\hline

Obs. & Mission & Sequence ID & Obs.\ Start (UTC) & Exp. (ks) \\
\hline
1 & NuSTAR & 30601007002 & 2020-04-15 16:36:09 & $31.0$\\
& NICER & 3604020101 & 2020-04-15 19:59:00 & $1.7$ \\

2 & NuSTAR & 30601007004 & 2021-04-18 06:01:09 & $  31.4$\\
& NICER & 3604020104 & 2021-04-18 06:13:52 & $  4.8$ \\

\end{tabular}
\end{center}
\end{table}

\begin{figure}[t!]
\begin{center}
\includegraphics[width=0.48\textwidth, trim = 18 0 23 0, clip]{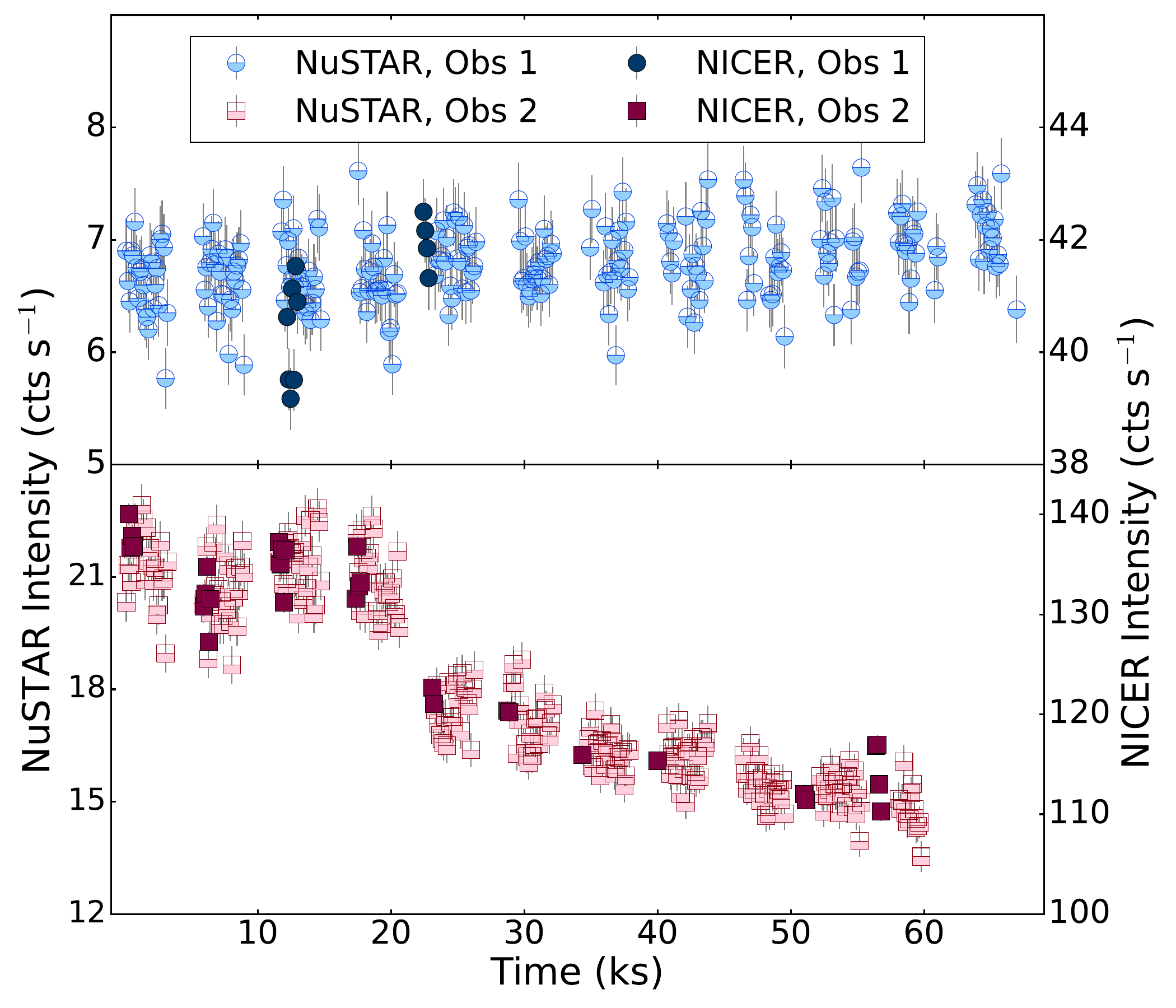}
\caption{Light curve for the \nustar (half-filled)  and \nicer\ (filled) observations of \source binned to 128~s. The top panel represents Obs 1 (blue circles) and the bottom represents Obs 2 (red squares). Only one NuSTAR focal plane module (FPM) is shown for clarity.}
\label{fig:lc}
\end{center}
\end{figure}

\begin{figure}[h!]
\begin{center}
\includegraphics[width=0.48\textwidth, trim = 22 0 23 0, clip]{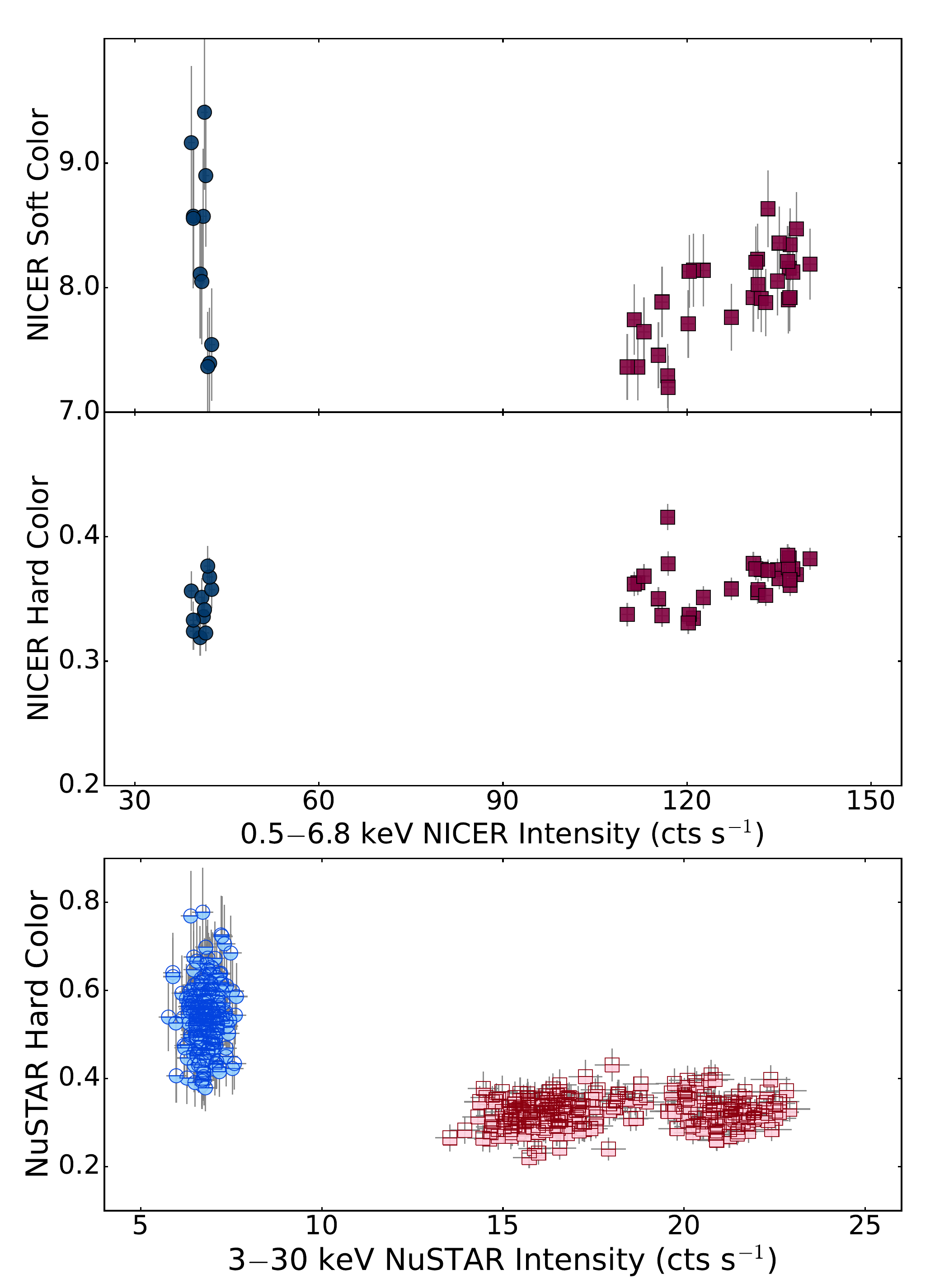}
\caption{Color-Intensity diagrams for \nicer and \nustar observations of \source. Markers match those used in Figure 2. The top and middle panels represent the soft and hard color in \nicer (respectively defined using the bands 1.1-2.0 keV/0.5-1.1 keV and 3.8-6.8 keV/2.0-3.8 keV), and the bottom panel represents the hardest color measurement in \nustar, defined using the bands 10-16 keV/6.4-10 keV.}
\label{fig:cid}
\end{center}
\end{figure}

Neither observation contained a Type-I X-ray burst, so no additional filtering was done. Both the \nicer and \nustar data are rebinned using the optimal binning method \citep{kaastra16} with the requirement that each bin contains at least 30 counts to allow for the use of $\chi^2$ statistics. Figure \ref{fig:lc} shows the \nicer and \nustar light curves for both observations. Figure \ref{fig:cid} shows the color intensity diagrams for these observations. It is evident that the lower flux observation corresponds to the source in a hard spectral state, whereas the second observation captured the source in a softer spectral state.

\begin{figure}[h!t]
\begin{center}
\includegraphics[width=0.48\textwidth, trim = 8 30 40 30, clip]{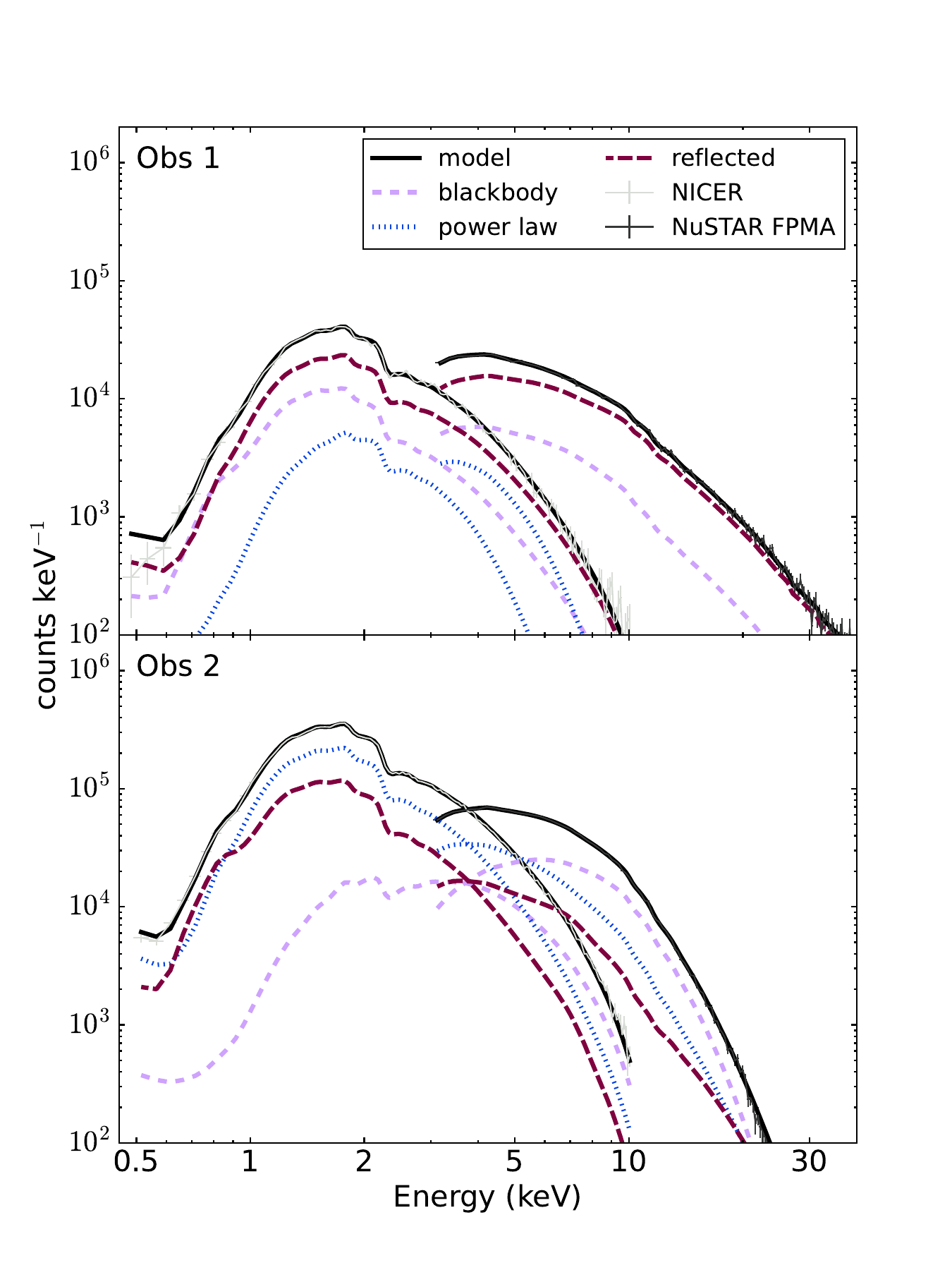}
\caption{The  \nicer and \nustar spectra in units counts keV$^{-1}$ and the respective model components from \xillverco for both observations. We can see here that Obs 1 has an overall lower flux and different shape to that of Obs 2, which displays a cutoff in the power law around 5.8 keV.} 

\label{fig:spec}
\end{center}
\end{figure}

\section{Spectral Modeling and Timing Analysis}\label{sec:results}
In this section we discuss the process used to model both the continuum and the reflected emission in the low flux, hard state (Obs 1) and higher flux, soft state (Obs 2), as well as an analysis of some of the timing properties of this system.  The \nicer spectra are presented in the band from 0.45 to 10 keV, while the \nustar spectra are in the 3-40 keV band. Certain regions of both spectra are background dominated. In Obs 1, the very lowest energies ($\lesssim0.7$ keV) encroach on the background, whereas in Obs 2 the highest energies are background dominated ($\gtrsim25$ keV), though it is source dominated all the way down to the lowest energies. This is consistent with Obs 1 being in a low hard state ad Obs 2 being in a high soft state.

\subsection{Continuum modeling}\label{subsec:cont}
 We begin by modeling the spectrum of \source with only a continuum description. This continuum is comprised of a blackbody of temperature, $kT_{\mathrm{bb}}$, representing the thermal emission from the NS, and a cutoff power law representing the illuminating corona with an index of $\Gamma_{\mathrm{pl}}$. We account for absorption of the continuum along the line of sight with {\sc tbabs} with a hydrogen column density of $N_H$ with {\sc wilm} abundance \citep{wilms00}. 

We reconcile the calibration differences between \nicer and \nustar using a model of the form $CE^{-\Delta\Gamma}$ \citep{steiner10}. We hold the constant $C$ to be 1 for the \nustar focal plane module A (FPMA) spectrum, and fix $\Delta\Gamma = 0$ in both \nustar spectra. We allow the constant to vary in the \nustar FPMB and in \nicer, and allow $\Delta\Gamma$ to vary in \nicer to adjust for the difference in slope due to calibration differences. 

With the model in place we fit the data for both observations using the {\sc xspec} v12.13.1 \citep{arnaud96}. With a reasonable starting place, we then run a Markov Chain Monte Carlo (MCMC) fit in {\sc xspec} with 100 walkers, a burn in of 100000 and a length of 10000.  The results of that fit are listed in Table \ref{tab:contTable}. We can see that notably the cutoff energy is much lower for Obs 2, indicating alongside Figure \ref{fig:cid} that the system entered a softer spectral state. The high-energy cutoff expected for LMXB systems frequently moves to lower energies \citep{degenaar18}. This cutoff at lower energies is visible in the shape of the spectra for Obs 2, which can be seen in Figure \ref{fig:spec}. Obs 1 smoothly follows a single power law within the bounds of these spectra, and so the cutoff energy for this observation should not be considered a physical result. The results of the power-law model in Obs 1 (the harder spectral state) are roughly consistent with \cite{david97}, who find that the continuum can be effectively modeled using an absorbed power law with an index of $\sim2$ (though we may expect the value to vary as the source changes between states).  In order to ensure that our results are not affected by the choice of background model, we also fit the continuum using the SCORPEON model\footnote{https://heasarc.gsfc.nasa.gov/docs/nicer/analysis\_threads/scorpeon-overview/}, and find that continuum parameters agree within uncertainty for Obs 2, and the fit values agree within uncertainty for all continuum parameters of Obs~1 except for the high-energy cut off, which again sits outside of the bands of the NuSTAR data, and hence can not be reliably constrained. Because of this, we opt to continue our analysis using the 3C50 background to minimize the number of free parameters. This is consistent with \cite{partington23}, which indicates that 3C50 is sufficient even down to source count rates of $\sim1$ count s$^{-1}$.

\begin{figure}[!t]
\begin{center}
\includegraphics[width=0.48\textwidth, trim = 20 30 0 0, clip]{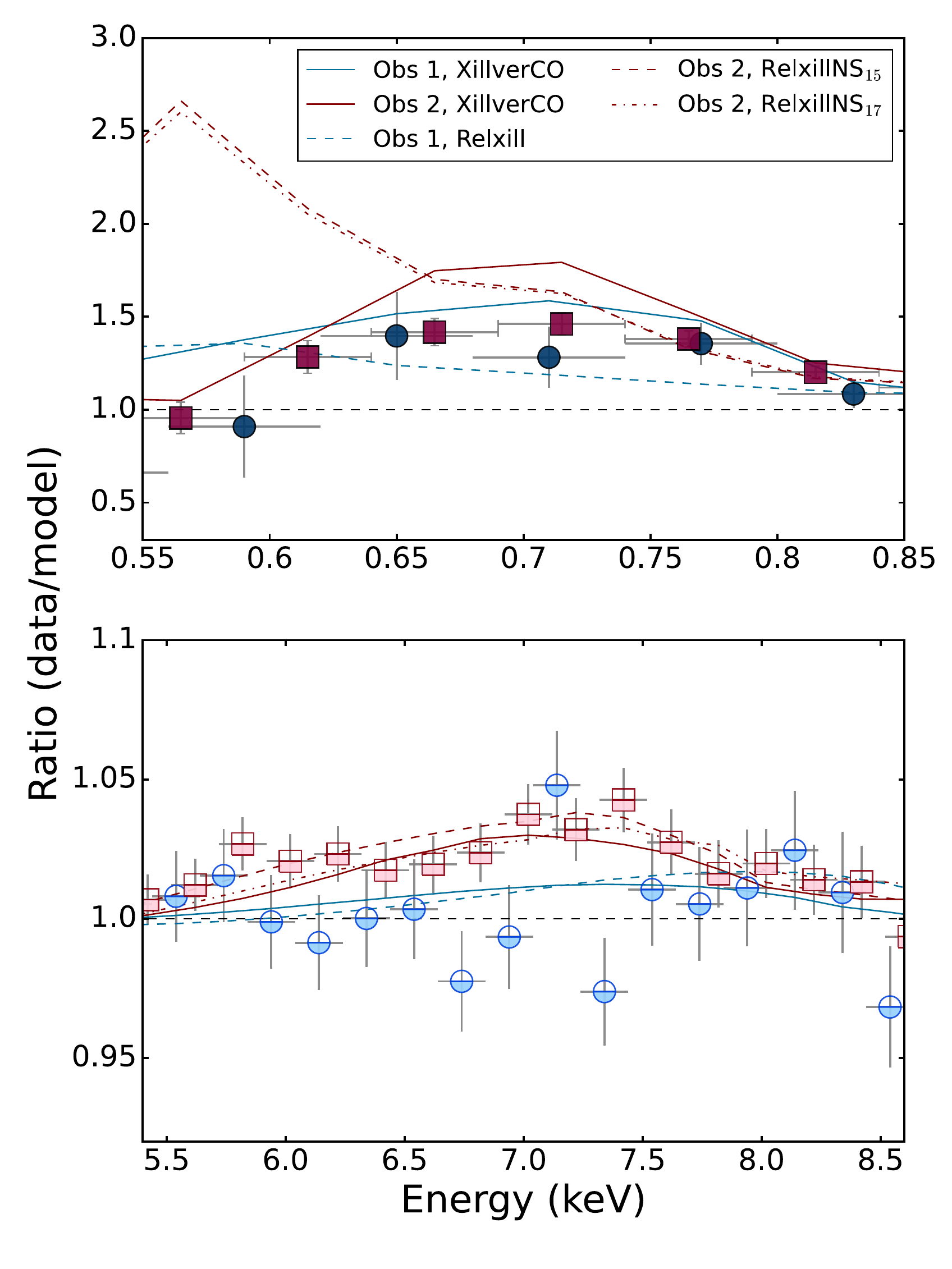}
\caption{Shown here are the regions containing the (top) \ion{O}{8} Ly$\alpha$ in \nicer and (bottom) Fe K$\alpha$ reflection in \nustar FPMA. These are constructed by ignoring the regions surrounding the line in {\sc xspec}, fitting a continuum, then reintroducing these regions and plotting the ratio of the data to the model. The subscripts 15 and 17 refer to the value at which logN is fixed for \relxillns in each model. We see strong evidence for a feature around the \ion{O}{8} energy band, but a feature around the Fe K$\alpha$ energy band is significantly weaker, at only about 3-4\% above continuum. We show the different models used and find that only \xillverco effectively detects the feature in the lower energy bands.}
\label{fig:lines}
\end{center}
\end{figure}

\begin{table}[]

\caption{Continuum}
\label{tab:contTable}
\begin{tabular}{lll}
& Obs 1          & Obs 2           \\
\hline\hline
C$_{FPMB}$ &$0.983\pm0.001$    & $1.001\pm0.004$      \\
C$_{NICER}$ & $0.85_{-0.08}^{+0.04}$       & $0.89\pm0.03$          \\
$\Delta\Gamma$ (10$^{-2}$) &$-7.0_{-6.7}^{+3.6}$ & $-8.6_{-1.8}^{+2.4}$       \\
$N_{\rm H}$ (10$^{22}$ cm$^{-2}$) &$1.62\pm0.04$ &$1.51\pm0.02$ \\
$kT_{\mathrm{bb}}$ (keV)&  $0.69_{-0.06}^{+0.04}$     &$2.01\pm0.02$ \\
k$_{\mathrm{bb}}$(10$^{-3}$) &$0.34_{-0.07}^{+0.04}$         & $3.0\pm0.1$       \\
$\Gamma_{\mathrm{pl}}$&$1.92_{-0.02}^{+0.07}$    &  $1.7\pm0.1$\\
E$_{\mathrm{cut, pl}}$ (keV) &$125_{-41}^{+123}$       &$5.9_{-0.3}^{+0.2}$\\
k$_{\mathrm{pl}}$ &$0.072_{-0.003}^{+0.009}$           &$0.25\pm0.01$ \\
\hline
$\chi^2$ (dof)             & 428(386) & 676(356)

\end{tabular}
\medskip

Note: All errors are reported at the 90\% confidence interval. The blackbody normalization (k$_{\mathrm{bb}}$) is defined as $(L/10^{39}$ erg s$^{-1})/(D/10$ kpc)$^2$, and the powerlaw normalization (k$_{\mathrm{pl}}$) is defined as photons keV$^{-1}$ cm$^{-2}$ s$^{-1}$ at 1 keV.
\end{table}

\begin{figure}[!t]
\begin{center}
\includegraphics[width=0.47\textwidth, trim = 50 130 0 0, clip]{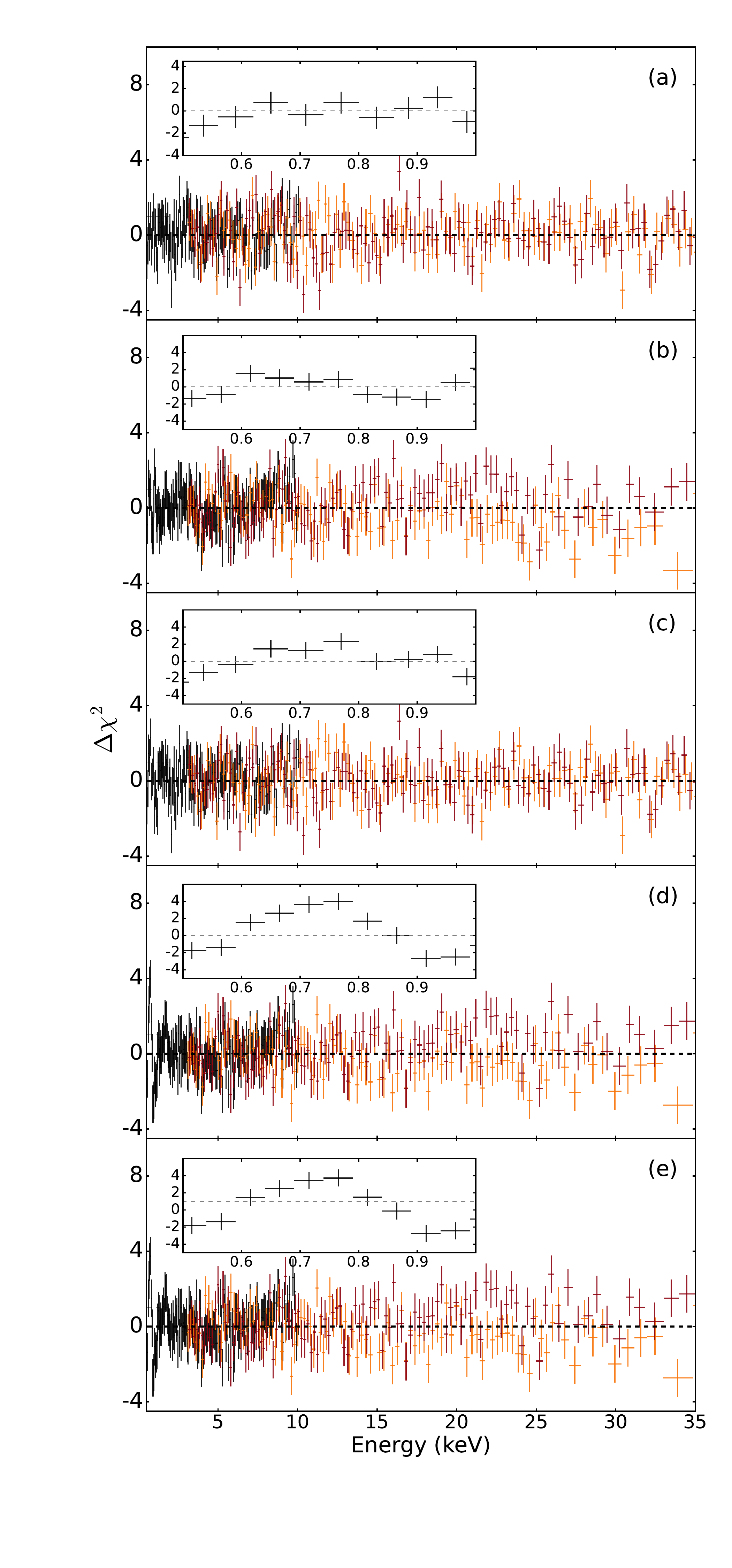}
\caption{Shown above are the model residuals for the full X-ray band in NICER (black) and NuSTAR (Orange/Red for FPMA/FPMB respectively). The panels show the residuals for (a) Obs 1 using \xillverco, (b) Obs 2 using \xillverco, (c) Obs 1 using \relxill, (d) Obs 2 using \relxillns with $\log N = 15$, and (e) Obs 2 using \relxillns with $\log N = 17$. The insets display the \nicer 0.5--1.0 keV energy range to highlight the difference in fit quality for the area in which we expect an \ion{O}{8} Ly$\alpha$ for an UCXB.}
\label{fig:residuals}
\end{center}
\end{figure}

We look for visual evidence of reflected features by inspecting regions surrounding the expected features. We initially ignore data bins between 0.6--0.8 keV (corresponding to the \ion{O}{8} Ly$\alpha$ feature expected for CO WD UCXBs at around 0.67 keV) and 5.5--7.4 keV (corresponding to the Fe K$\alpha$ feature around 6.4 keV). We fit the continuum with these regions ignored, then reintroduce them and plot the ratio of the data to the model. The results of this plotting can be seen in Figure \ref{fig:lines}.  We see that the feature around the energy band of Fe K$\alpha$ peaks at around 4\%, quite a bit lower than the 10-15\% seen in some canonical LMXBs (for example, Ser X-1, see \citealt{ludlam18}). However, a very strong feature (around 50\% above continuum) is seen at the lowest energies. It should be noted that these features can often be overestimated if other absorption effects such as absorption edges are not accounted for \citep{ludlam21}. We find, however, that edges included in the model are poorly constrained and do not significantly impact fit quality, so we exclude these from our model. 

To be confident in the existence of these reflection features, we fit our continuum model with two additional Gaussian components. These components have their central energy fixed at 0.67 keV and 6.4 keV, to account for \ion{O}{8} and Fe K$\alpha$ respectively. These Gaussians improve the fit quality in Obs 1 by $>5.5\sigma$ and Obs 2 by $>8.9\sigma$ via F-test, with equivalent widths (EWs) that are often consistent with other reflection features seen in X-ray binaries. For example in Obs 1, the EW is 165 eV for the \ion{O}{8} feature, which sits on the high end of other measured EWs (see \citealt{cackett10} for examples of EWs of Fe K$\alpha$ lines in LMXBs, and \citealt{madej11} for examples of EW in UCXBs). The Fe K$\alpha$ feature has a high EW value, at around 400 eV. This is reflective of the low contribution of the feature, and indicates that a gaussian does not effectively pick up any prominent features around 6.4 keV. Similarly, in Obs 2, the EWs for \ion{O}{8} and Fe K$\alpha$ are 110 eV and 326 eV, respectively. The values for these Gaussian parameters including their EW and normalization can be found in Appendix A.

To ensure that our detection of emission lines indicative of reprocessed emission does not hinge upon our choice of continuum model, we check their presence with the use of two additional continuum model descriptions utilized for NS LMXBs. These models are an absorbed blackbody with thermal comptonization ({\sc tbabs*(nthcomp + blackbody)} with {\sc nthcomp inp\_type} set to 0 for a blackbody input) and an absorbed disk with thermal comptonization ({\sc tbabs*(nthcomp + diskbb)} with {\sc nthcomp inp\_type} set to 1 for a disk blackbody input). {\sc nthcomp} is a model which replaces the continuum component often modeled as a simple powerlaw with a more physically motivated thermally comptonized plasma \citep{zdziarski96, zycki99}. In each case we find a $> 5 \sigma$ improvement to the fit when Gaussian features are included. This is consistent with other similar analyses, which indicate that the continuum description does not impact the detection of line features \citep{coughenour18,ludlam20,ludlam22}. The alternative continuum descriptions also display relatively narrow \ion{O}{8} features, with poorly constrained Fe K$\alpha$ equivalent widths. The results of these models can also be found in Appendix A. Given the robust detection of the emission lines regardless of continuum model, we proceed with modeling the reprocessed emission with our primary continuum description given the availability of self-consistent reflection models.

\subsection{Reflection Modeling}\label{subsec:refl}
As mentioned previously, the reprocessed emission from an externally illuminated accretion disk contains information about the physical properties of the emitting material in the region close to the compact object, and therefore can be utilized to learn about the properties of the accretor and disk (e.g., chemical composition, ionization state, system inclination, etc.; \citealt{ludlam24}). For testing the chemical composition of the accretion disk, we utilize reflection models that differ significantly in chemical abundance.

First, we apply \xillverco which is a reflection table based on \xillver \citep{garcia10,garcia13} with carbon and oxygen abundances similar to what is seen in CO WDs (i.e., the disk is nearly devoid of H and He while overabundant in C and O). This reflection table has been used in the literature to model the reprocessed emission spectra of several UCXBs \citep{madej14, ludlam21, moutard23}. The model produces the reprocessed spectrum assuming primary illumination by a cutoff powerlaw and contains the emergent blackbody component at the emission radius of reflection. We allow the CO abundance (A$_{\mathrm{CO}}$), the disk temperature at the region of reflection (kT$_{\mathrm{refl}}$), the ratio of the incident flux to that of the emergent blackbody flux at the region of reflection (`frac'), and the normalization to be free\footnote{See \cite{dauser16} and \cite{moutard23} for more discussion of the normalization of \xillverco and \relxillns.}. We tie the E$_{cut}$ in \xillverco to that of the cutoff powerlaw. We fix the redshift to 0 since the source is Galactic.

\begin{table*}[t!]
\begin{center}
\caption{Reflection Model Comparison}
\label{tab:allmodels}
\begin{tabular}{lccc|cc}

& Relxill &\multicolumn{2}{c}{RelxillNS} &\multicolumn{2}{c}{XillverCO}  \\
\hline
& Obs 1          & \multicolumn{2}{c}{Obs 2}          & Obs 1          & Obs 2         \\     
\hline
C$_{FPMB}$ &$0.99\pm0.01$   &$1.001_{-0.005}^{+0.004}$ & $1.001\pm0.003$  &$0.983_{-0.005}^{+0.002}$ &$1.000_{-0.002}^{+0.006}$\\
C$_{NICER}$ &$0.88_{-0.02}^{+0.01}$  &$0.83\pm0.02$   &$0.83\pm0.01$  &$0.87_{-0.03}^{+0.05}$ &$0.85_{-0.04}^{+0.02}$\\
$\Delta\Gamma$(10$^{-2}$) & $-3.9_{-1.0}^{+0.7}$ &$-12.5_{-1.7}^{+1.4}$  &$-12.9\pm0.1$  &$-5.1_{-2.5}^{+4.4}$ &$-11.4_{-2.9}^{+1.5}$ \\
$N_{\rm H}$ (10$^{22}$ cm$^{-2}$) & $1.55_{-0.04}^{+0.01}$ &$1.32_{-0.01}^{+0.02}$ &$1.27\pm0.01$ &$1.67_{-0.03}^{+0.05}$ &$1.66_{-0.02}^{+0.04}$\\
kT$_{\mathrm{bb}}$ (keV)&$0.67_{-0.02}^{+0.04}$ &$1.67_{-0.04}^{+0.02}$    &$1.60_{-0.03}^{+0.04}$    &$0.67\pm0.03$ &$2.05_{-0.01}^{+0.03}$   \\
k$_{\mathrm{bb}}$(10$^{-3}$) & $0.52_{-0.03}^{+0.04}$ &$3.4\pm0.2$       &$3.67\pm0.05$ & $0.37_{-0.05}^{+0.03}$ & $3.5_{-0.2}^{+0.1}$ \\
$\Gamma_{\mathrm{pl}}$&$1.77\pm0.01$ &$0.53_{-0.05}^{+0.03}$   &$0.149\pm0.005$ &$1.93_{-0.03}^{+0.04}$ &$1.69_{-0.024}^{+0.16}$\\
E$_{\mathrm{cut, pl}}$ & $72.7_{-3.3}^{+6.2}$ &$1.32_{-0.05}^{+0.02}$ & $1.05\pm0.02$ & $145_{-36}^{+40}$ &$5.0_{-0.2}^{+0.6}$  \\
k$_{\mathrm{pl}}$ &$0.036_{-0.002}^{+0.003}$   &$0.33\pm0.01$  &$0.36_{-0.02}^{+0.01}$     &$0.07\pm0.01$ &$0.17_{-0.01}^{+0.07}$\\
\hline
q      &$4.8_{-1.5}^{+1.1}$     &$10^\dagger$    &$9.98_{-0.13}^{+0.02}$     &$3.6_{-1.4}^{+0.9}$ &$2.5_{-0.5}^{+2.5}$  \\
$i$ (deg) & $69.4_{-3.7}^{+3.6}$ & $50.1_{-4.2}^{+2.8}$ & $46.5_{-1.3}^{+2.1}$     &$59.4_{-3.1}^{+21.2}$ &$54.9_{-4.7}^{+19.0}$\\
$R_{in}$ (R$_\mathrm{ISCO}$) &$1.3_{-0.2}^{+0.1}$  &$1.8\pm0.2$     &$1.51_{-0.04}^{+0.03}$     & $1.4_{-0.4}^{+2.5}$ &$1.7_{-0.6}^{+5.8}$   \\
A$_{\mathrm{CO}}$ &--- &--- &---  &$47_{-11}^{+40}$ &$38_{-3}^{+16}$            \\
A$_{\mathrm{Fe}}$ &$1.9_{-0.5}^{+0.2}$ & $0.51_{-0.01}^{+0.08}$ &$0.5^\dagger$  &--- &---          \\
kT$_{\mathrm{\xillverco}}$ (10$^{-2}$ keV) &---     &--- &---  & $8.7_{-2.1}^{+0.1}$ &  $9.6_{-0.8}^{+0.3}$      \\
kT$_{\mathrm{\relxillns}}$ &---     &$2.8\pm0.1$ &$2.90\pm0.04$  & --- & ---      \\
frac &--- &---  &---  &$0.11_{-0.02}^{+0.11}$&$0.10\pm0.02$\\
$|\mathrm{f}_{\mathrm{refl}}|$ &$0.9\pm0.1$ & $5.1_{-0.4}^{+0.3}$  &$5.2_{-0.1}^{+0.2}$  &---&---\\
k$_{\mathrm{\xillverco}}$ (10$^{-9}$) & ---     &---     &---  &$0.5_{-0.1}^{+0.5}$ & $3.1_{-0.4}^{+1.2}$  \\
k$_{\mathrm{\relxill}}$ (10$^{-4}$) & $5.9_{-0.6}^{+0.3}$    &$0.7\pm0.1$    &$0.70\pm0.01$  &--- &---  \\

$\log{N} ({\mathrm cm}^{-3})$ &---          &15$^*$ &17$^*$           &\multicolumn{2}{c}{---}  \\
$\log{\xi}$  & $3.8_{-0.2}^{+0.1}$    & $3.5\pm0.1$  &$3.6\pm0.1$    &$2.9\pm0.5^A$ & $3.0\pm0.1^A$      \\

\hline
$F_{\mathrm{2-10}}$ (10$^{-10}$ erg s$^{-1}$ cm$^{-2}$) & $1.93\pm0.01$ & $5.85\pm0.01$ & $5.86\pm0.01$ &$1.93\pm0.01$ &$5.84\pm0.01$\\
$F_{\mathrm{0.5-50}}$ (10$^{-10}$ erg s$^{-1}$ cm$^{-2}$) & $5.73\pm0.02$ & $11.26\pm0.02$ & $11.01\pm0.02$ & $6.02\pm0.02$ &$ 15.98\pm0.02$\\
F$_\mathrm{Edd}$ (10$^{-2}$) & $0.61\pm0.01$ & $1.19\pm0.01$ & $1.17\pm0.01$ & $0.64\pm0.01$ & $1.69\pm0.01$\\
\hline
$\chi^2$ (dof)             & 400 (379) & 439 (348) & 431(348) & 395(378) & 415 (349)\\

\end{tabular}

\medskip
Note: $^*=$ parameter is fixed. $^A=$ not a model component and is calculated using the description in Section \ref{subsec:refl}, using the largest errors for a conservative estimate. All errors are reported at the 90\% confidence interval. For comparison some rows are used for both \relxillns and \xillverco despite having slightly different definitions in their respective models. A$_{\mathrm{CO}}$ refers to the carbon and oxygen abundance in \xillverco and the iron abundance (A$_{\mathrm{Fe}}$) refers to the Fe abundance in \relxill and \relxillns. frac represents the ratio of the illuminating powerlaw to that of the emergent blackbody from the disk (kT$_{\mathrm{refl}}$) in \xillverco, whereas f$_{\mathrm{refl}}$ represents the ratio of the illuminating X-rays to those that escape to infinity in both \relxill and \relxillns. The blackbody normalization (k$_{\mathrm{bb}}$) is defined as $(L/10^{39}$ erg s$^{-1})/(D/10$ kpc)$^2$, and the powerlaw normalization (k$_{\mathrm{pl}}$) is defined as photons keV$^{-1}$ cm$^{-2}$ s$^{-1}$ at 1 keV. The normalization for \relxill and \relxillns k$_{\mathrm{\relxill}}$ scales differently than that of \xillverco k$_{\mathrm{\xillverco}}$ so we separate these for ease of reading. $F_{\mathrm{0.5-50}}$ refers to the unabsorbed flux in the band 0.5-50 keV, where $F_{\mathrm{2-10}}$ refers to the observed flux in the 2-10 keV band. F$_\mathrm{Edd}$ refers to the Eddington ratio, calculated using the 0.5-50 keV flux and the empirical Eddington luminosity of 3.8$\times10^{38}$~\lumcgs.
\end{center}
\end{table*}

Because \xillverco does not account for the relativistic broadening of reflected features, we must convolve the model with {\sc relconv}, which can be used to determine certain physical parameters of the system \citep{dauser10}. Specifically we can use it to determine the inner disk radius at which the reflection is occurring (R$_{in}$), which we report in terms of the innermost stable circular orbit (\risco), the orbit at which a test particle can orbit stably without falling onto the neutron star, 6 gravitational radii \rg=12.4 km for a 1.4 M$_\odot$, non-rotating neutron star), assuming the reflection occurs at the innermost region of the disk. In this model we assume that for a NS only one emissivity index is necessary, so we fix both indices to be equal $q_1=q_2=q$ and fix the break radius to 500\rg\ (an obsolete parameter given that there is a single emissivity index). Since the X-rays only probe the innermost region of the disk, we fix the outer disk to 990 \rg. We tie the inclination measured from {\sc relconv} to that measured with \xillverco. We also fix the limb parameter and the dimensionless spin to 0. We choose 0 following \cite{ludlam18}, who shows that the effect of the spin is minimal for most LMXBs. The fitting process described in Section \ref{subsec:cont} is repeated, and the results are shown in the rightmost two columns of Table \ref{tab:allmodels}. 

In order to compare the results of a model with UCXB abundances to those with standard solar abundances typical of a standard NS LMXB, we then replace the {\sc relconv*xillverCO} in Obs 1 and Obs 2 with \relxill \citep{garcia14} and \relxillns \citep{garcia22} , respectively, to account for relativistic reflection. The difference between these models lies in the illuminating source-- \relxillns assumes the disk is illuminated by a thermal component such as the boundary layer or hot spot on the NS, whereas \relxill uses a cutoff powerlaw to describe illumination by a hot electron corona. The different spectral states between our two observations necessitates the use of different models accordingly when testing for a disk composed of solar abundance material. Many of the parameters remain the same between all three models, with a few exceptions.For one, in \relxill and \relxillns, the abundance of all elements in the accretion disk are set to solar values with the exception of a variable iron abundance (A$_{\mathrm{Fe}}$).

The reflection fraction f$_{\mathrm{refl}}$ parameter in Table \ref{tab:allmodels} represents the ratio of illuminating flux to that which is reflected for \relxill and \relxillns. This is fixed to negative values during fitting in order to only model the reflected component, but the absolute value is reported in Table \ref{tab:allmodels}. The ionization state of the material is given by $\log\xi$ (where $\xi = \frac{4\pi}{n} F_x$). \relxillns has a variable disk density component ($\log N [\rm cm^{-3}]$) that varies from 15 -- 19. We note that both \relxill and \xillverco have fixed disk density of $\log N [\rm cm^{-3}] = 15$ and $\log N [\rm cm^{-3}]= 17$, respectively. To properly compare \relxillns to the other models, we perform fits with the density fixed at both 15 (the fixed value in \relxill; \citealt{garcia14}) and 17 (the fixed value in \xillverco; \citealt{madej14}) . The results of these models are found in Table \ref{tab:allmodels}. Since $\log\xi$ is not a parameter in \xillverco, we calculated it using the definition above, where $F_x = {\rm frac} \times \sigma T^4$, with the value of temperature T from kT$_{\mathrm{refl}}$ in \xillverco (see \citealt{ludlam21} for more information). We use the multiplicative model {\sc cflux} in {\sc xspec} to calculate the unabsorbed fluxes in the 0.5--50 keV band, as well as the absorbed (i.e. measured) flux in the 2--10 keV band. We then use the unabsorbed flux in conjunction with a recent distance measurement from \cite{galloway20} to measure the luminosity, $d = 5.8\pm0.9$ kpc \footnote{\cite{galloway20} poses two possible distances, we opt for the higher of the two to be closer to previously inferred distances of 8.5 kpc \citep{david97}}. We compare this to the empirical Eddington luminosity for a 1.4 M$_{\odot}$ NS $L_{Edd} = 3.8\times10^{38}$~\lumcgs \citep{kuulkers03} to calculate the eddington ratio F$_\mathrm{Edd}$. Both fluxes and F$_\mathrm{Edd}$ are reported in Table \ref{tab:allmodels}.

Figure \ref{fig:residuals} shows visually that the non-UCXB models perform worse in the lower energy than \xillverco. Because Obs 2 is background dominated above roughly 30 keV, the residuals are dominated by this regime, but we note that \relxillns also under performs in this regime. Regardless, \xillverco performs about equally well in energy ranges outside of these. We can also see from Figure \ref{fig:lines} that the general shape of the low energy feature is more closely followed by \xillverco than either \relxill or \relxillns.

\begin{figure}[h!t]
\begin{center}
\includegraphics[width=0.48\textwidth, trim = 8 30 40 30, clip]{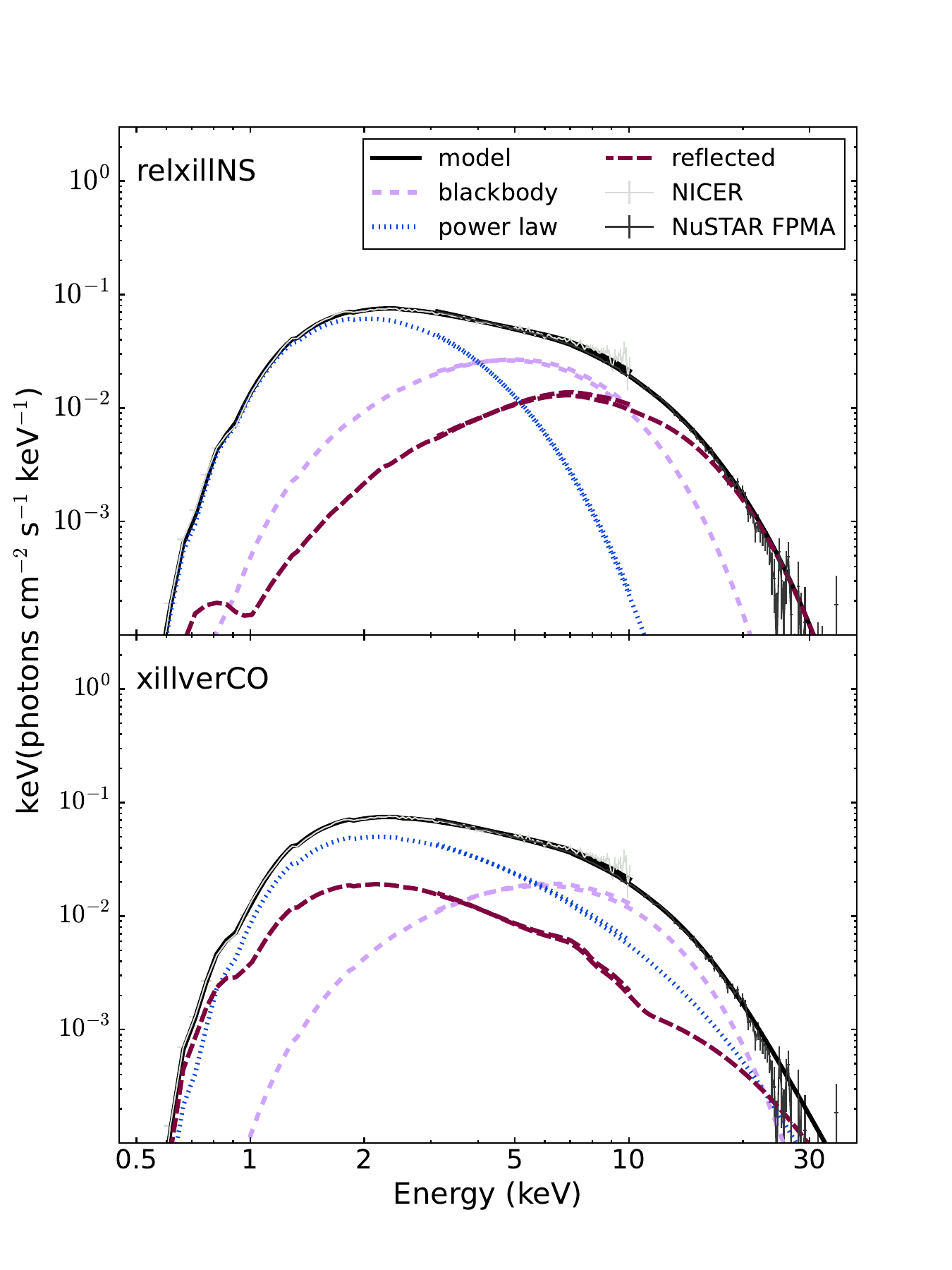}
\caption{This plot compares the unfolded spectrum for Obs 2 when modeled with \relxillns with logN fixed at 17 to the spectrum modeled with \xillverco (the same spectrum shown in Figure \ref{fig:spec}). We see that the reflection component shifts strongly toward higher energies, and so the powerlaw and blackbody must shift toward lower energies to account for this.}
\label{fig:obs2comp}
\end{center}
\end{figure}

The differences in certain continuum parameters between the models can be explained by continuum model components compensating for a low energy feature that is not present in the canonical LMXB models. This can be seen in Figure \ref{fig:obs2comp}, which demonstrates how the powerlaw and blackbody components change to adjust to a reflection model which does not encompass the low energy features. \relxill and \relxillns are both designed with relativstic broadening. Therefore, in order to compare \xillverco (which has no broadening inherent to the model) with these models, we must verify that the features detected are also relativistically broadened. We test the spectral broadening by fitting the data both with and without {\sc relconv}, and we find that in Obs 1, the $\chi^2$ improves by 11 for 2 degrees of freedom with the addition of {\sc relconv}, and in Obs 2 the $\chi^2$ improves by 28 for 2 degrees of freedom. This implies that the features detected by \xillverco are broadened with some degree of significance. 

\subsection{Timing Analysis}\label{subsec:timing}
Since the key defining parameter of a UCXB is a period of $<$ 80 minutes, we attempt to search for evidence of periodicity in the X-ray light curves. \cite{wijnands99} suggests that the inclination angle of the source may be quite high, leading to a smearing of pulsations. A high inclination should result in eclipses in the X-ray light curve, yet none are seen in the data, which is supported by \cite{intzand07}.  We search \nustar light curves for evidence of periodicity. These light curves were barycenter corrected using the {\sc barycorr} tool in {\sc heasoft}. A search for the presence of periodic signals was performed on the light curves from each of the two individual observations using the Z$^2$ test \citep{buccheri83}. We employed the Z search algorithm in Stingray \citep{huppenkothen19} to perform the search in a grid of frequencies corresponding to periods between 10 minutes to 90 minutes, which is physically motivated based on the expected orbital period range for UCXBs. The search resulted in the detection of peaks ($>$ 3 $\sigma$) only around the harmonics of the NuSTAR orbital period of 96.8 minutes. Based on our analysis we estimate an upper limit on the amplitude of periodic signals in the mentioned frequency range to be 5--6\% with a 99\% confidence limit. Further analysis with Fourier methods also yields no significant period measurement. This is, however, unsurprising, as the short exposures and low count rates of \source are unlikely to provide strong constraints on any sort of timing analysis. For further discussion of the timing properties see \cite{wijnands99}, who find that most timing properties of this source are consistent with other NS LMXBs. The exception to this behavior is the power spectrum break frequency, which is anti-correlated with the X-ray flux; this trend is reversed in most NS LMXBs.

\section{Discussion and Conclusions}\label{sec:discussion}
Between the two observations of \source, we find that the spectral shape changes significantly from a powerlaw dominated continuum to one dominated by thermal emission with a lower energy cut off. We find in Section \ref{sec:results} that the best fit statistics are achieved using \xillverco. Aside from the best fit statistics, the values retrieved from the fit are also generally more consistent with what we expect from a UCXB. Many of the reported parameters from \relxillns in Table \ref{tab:allmodels} are unrealistic. For example in Obs 2, the power law index is lower than 1, which is lower than the extreme hard spectral state for NSs \citep{ludlam16,parikh17}. A$_{\mathrm{Fe}}$ in both trials with \relxillns is also consistent with the lower bound of the model at 0.5. We also see that in both \relxill and \relxillns, the emissivity index q is unphysically high, approaching the upper bound of 10 for \relxillns. We attempt to freeze these values at $q=3$, a more reasonable value for NSs \citep{wilkins18, ludlam24}, but this only serves to worsen the $\chi^2$ further. While this should not be taken as direct support that \source is in fact a UCXB, this demonstrates that models with higher carbon and oxygen abundances do in fact provide a better explanation for the properties of the X-ray spectrum.

\begin{figure}[!t]
\begin{center}
\includegraphics[width=0.48\textwidth, trim = 15 10 0 0, clip]{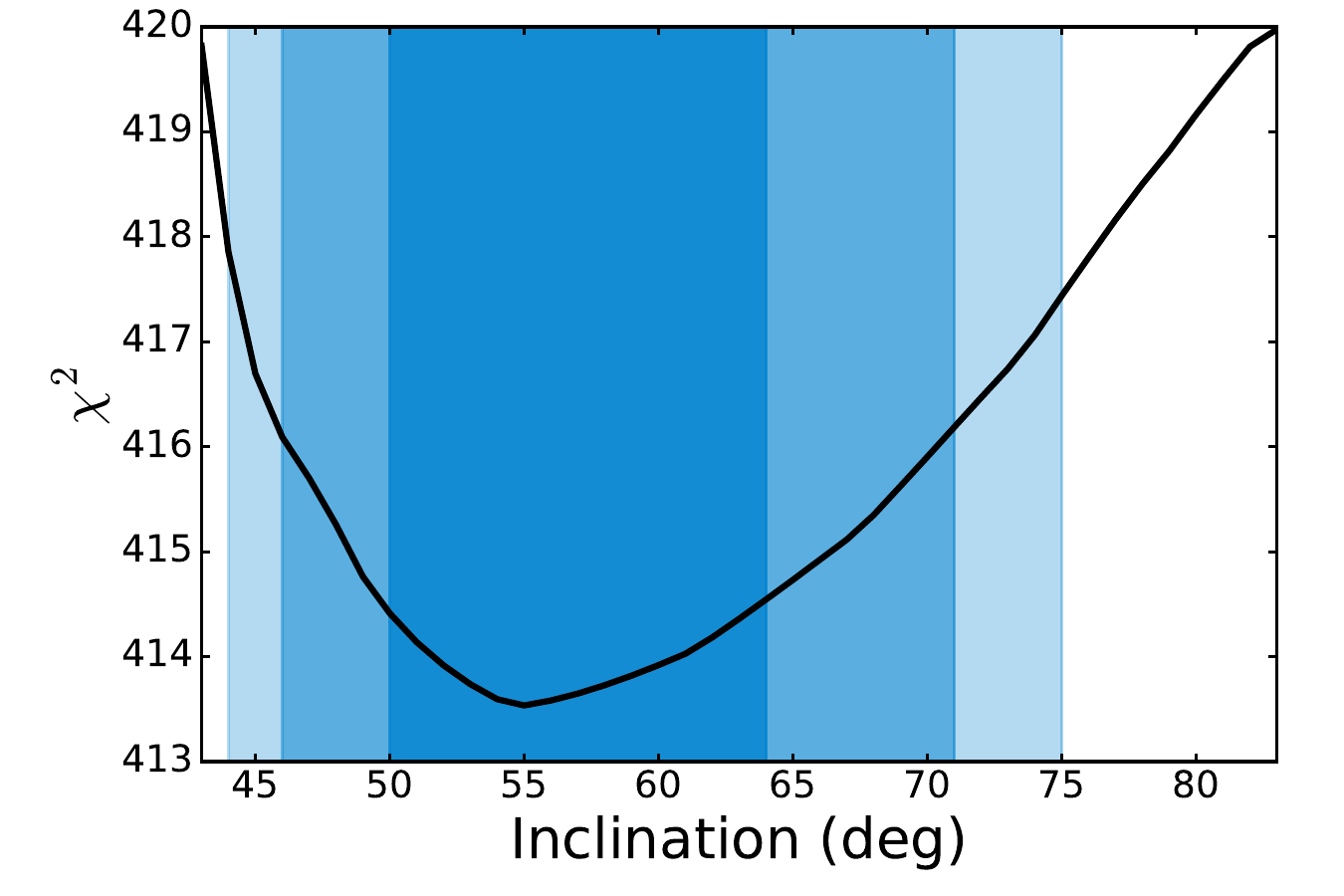}
\caption{We use the {\sc steppar} command in {\sc xspec} to determine the sensitivity and quality of fit for the inclination measurement. This steps the best {\sc xspec} fit through values of inclination while allowing all other free parameters to vary. The results may vary from those reported in \ref{tab:allmodels} which utilizes the results of MCMC. However, these 1-dimensional contours are more conservative limits given the {\sc xspec steppar} routine. We find that an inclination of around 55$^\circ$ is a global minimum, though there is a slightly stronger bias against low inclinations than higher ones. Shown here are the results of {\sc steppar} for Obs 2. The shaded regions represent the 68\%, 90\%, and 95\% confidence intervals. }
\label{fig:incl}
\end{center}
\end{figure}

It should be noted that the count rates of these spectra are relatively low, especially so for Obs 1, which could affect the quality of the reflection features. Because of this, certain parameters that may be of key interest, such as \rin, may not have the most reliable measurements. The \rin measurements listed in the \xillverco portion of Table \ref{tab:allmodels} indicate that we observe some minor disk truncation during the higher flux state. We generally expect the disk to move inward at higher luminosities, but magnetic fields can complicate this by truncating accretion disks even at higher luminosities \citep{cackett09, ludlam19}. It should be noted however that the existence of a feature surrounding the \ion{O}{8} Ly$\alpha$ energy range and none surrounding Fe K$\alpha$ prior to any modeling of reflection provides some evidence for non-solar carbon and oxygen abundances. This is also noted by \cite{koliopanos21}, who finds that screening by the C and O abundances leads to a diminished Fe K$\alpha$ feature in the UCXB sources 4U~1543$-$624 and Swift~J1756.9$−$2508.

We also present in this paper the first tentative measurement of the inclination of \source at approximately $57^{+23}_{-7}$ degrees by taking the mean of the \xillverco measurements and using the maximum upper and lower bounds of both observations to define the uncertainty. As mentioned above, we do not expect an especially high inclination as we do not observe eclipsing in the light curve. We test whether this parameter is a significant contributor to the statistics in the fit using the {\sc steppar} command in {\sc xspec}, which varies just one parameter and measures the change in $\chi^2$. We find that there is some degree of sensitivity, with the $\chi^2$ value increasing by approximately 1 at 5 degrees on either side of the measured value, as shown in Figure \ref{fig:incl}.

After comparing multiple models and attempting various types of analysis we suggest the following:
\begin{enumerate}
    \item \xillverco appears to provide a better fit for the X-ray spectra of \source than \relxill or \relxillns, which indicates that the carbon and oxygen abundance deviates from solar. This,  alongside the evidence for a lower-energy oxygen feature, suggests that the source is more likely to be a UCXB than a canonical LMXB, strengthening the classification as a UCXB candidate. This is further supported by the fact that only \xillverco was able to model the feature in the low energy band. Relativistic broadening is statistically required to model these features, indicating they are in fact coming from reflection off of the rotating disk.
    \item The reflection modeling using \xillverco has provided a tentative measurement of the inclination of this system at $57^{+23}_{-7}$ degrees. The reflection features used to measure both \rin and inclination are not very prominent, so further observations with longer exposures are needed to confirm.
    \item Our timing analysis is inconclusive. The fact that no eclipses are present in the light curve is consistent with an inclination $\lesssim80^\mathrm{o}$. Our Fourier analysis of the system does not reveal any measurable periodicity in the X-ray light curve. This means we can not conclusively deem the source to be a UCXB by any timing periodicities. 
\end{enumerate}

This study provides some additional indirect evidence in support of \source being a UCXB candidate. We also present an early measurement of the inclination. Future studies using longer exposures in the soft X-rays (for example, from \nicer) will be necessary to measure the orbital period and determine whether the source is ultra-compact in nature. Optical spectral follow up could potentially provide useful data on the abundance of carbon and oxygen present in the spectrum. Recent missions like XRISM would be useful in resolving reflection features, especially for testing for the existence of the faint Fe K$\alpha$ in these ultra-compact systems \citep{gandhi22}. 
\\
\\
{\it Acknowledgements:} 
This research has made use of MAXI data provided by RIKEN, JAXA and the MAXI team \citep{matsuoka09}. This research has made use of data and/or software provided by the High Energy Astrophysics Science Archive Research Center (HEASARC), which is a service of the Astrophysics Science Division at NASA/GSFC. This research has made use of the \nustar Data Analysis Software (NuSTARDAS) jointly developed by the ASI Science Data Center (ASDC, Italy) and the California Institute of Technology (USA).

\appendix
\section{Testing the presence of emission lines by the addition of  Gaussians to different continuum model descriptions}
\begin{table*}[h!]
\caption{Results of Gaussian Fits}
\begin{tabular}{lll|ll|ll}
\label{tab:gauss}

& \multicolumn{2}{l}{\sc tbabs*(cutoffpl+bbody)} & \multicolumn{2}{l}{\sc tbabs*(nthcomp[0]+bbody)} & \multicolumn{2}{l}{\sc tbabs*(nthcomp[1]+diskbb)} \\
& Obs 1 & Obs 2 &Obs 1 & Obs 2&Obs 1 & Obs 2 \\

\hline\hline
E$_{\mathrm{O}}$ (keV) & 0.67$^\dagger$ &0.67$^\dagger$&0.67$^\dagger$&0.67$^\dagger$&0.67$^\dagger$&0.67$^\dagger$\\
$\sigma_{\mathrm{O}}$($10^{-1}$) (keV) &$8_{-1}^{+2}$ &$12\pm1$ &$1\pm0.04$ &$9_{-1}^{+3}$ &$1.0_{-0.4}^{+0.5}$ &$0.8\pm0.2$ \\
k$_{\mathrm{O}}$ (10$^{-2}$) &$1.3\pm0.8$ &$2.3_{-0.7}^{+0.8}$ &$1.5\pm0.1$ &$3.4_{-2.0}^{+3.0}$ &$2.4_{-1.1}^{+1.4}$ &$2.0_{-0.7}^{+0.6}$ \\
EW$_{\mathrm{O}}$ (eV) &165 &110 &67 &171 &59 &96 \\
\hline
E$_{\mathrm{Fe}}$ (keV) & 6.4$^\dagger$ &6.4$^\dagger$&6.4$^\dagger$&6.4$^\dagger$&6.4$^\dagger$&6.4$^\dagger$\\
$\sigma_{\mathrm{Fe}}$ &$2.0\pm0.3$ &$1.8\pm0.2$ &$2.0\pm0.3$ &$1.8_{-0.3}^{+0.2}$ &$2.3\pm0.4$ &$1.0\pm0.2$ \\
k$_{\mathrm{Fe}}$ (10$^{-3}$) &$0.7\pm0.3$ &$2.0_{-0.6}^{+1.6}$ &$0.5\pm0.2$ &$1.5^{+0.6}_{-0.5}$ &$0.6\pm0.3$ &$0.5\pm0.2$ \\
EW$_{\mathrm{Fe}}$ (eV) &405 &326 &262 &245 &303 &75 \\
\hline
Significance ($\sigma$) &$>5.6$ &$>8.9$ &$>5.1$ &$>8.9$ &$>5.0$ &$>5.2$\\

\end{tabular}

\medskip

Note: $\dagger$ indicates fixed value, same for all fits. Parameters with a subscript O refer to those measured around the expected \ion{O}{8} feature and those with the subscript Fe refer to the Fe K$\alpha$ feature. The normalization k is in units photons cm$^{-2}$ s$^{-1}$. EW is the equivalent width of the feature, measured using the {\sc eqwidth} command in {\sc xspec}. Significance refers to the improvement over the respective continuum model via f-test. 
\end{table*}
\end{document}